\documentclass[usenatbib]{mnras}

\usepackage{multirow}
\usepackage{rotating}
\usepackage{colortbl}
\usepackage{color}
\usepackage[fleqn]{amsmath}
\usepackage{amssymb}
\usepackage{amsfonts}
\usepackage{verbatim}
\usepackage{scalefnt}
\usepackage[percent]{overpic}
\usepackage[T1]{fontenc} 
\usepackage{aecompl} 
\usepackage{ulem}
\usepackage{cleveref}

\usepackage{float}

\def\lsim{\mathrel{\rlap{\lower 3pt \hbox{$\sim$}} \raise 2.0pt \hbox{$<$}}}
\def\gsim{\mathrel{\rlap{\lower 3pt \hbox{$\sim$}} \raise 2.0pt \hbox{$>$}}}

\newcommand{\comments}[1]{} 

\title[Improved gravitational radiation time-scales]{Improved gravitational radiation time-scales: significance for LISA and LIGO-Virgo sources}

\author[L. Zwick et al.]{Lorenz Zwick,$^{1,2}$\thanks{E-mail: zwicklo@ics.uzh.ch}
Pedro~R. Capelo,$^{1}$
Elisa Bortolas,$^{1}$\newauthor
Lucio Mayer$^{1}$ and
Pau Amaro-Seoane$^{3,4,5,6}$
\\
$^{1}$Center for Theoretical Astrophysics and Cosmology, Institute for Computational Science, University of Zurich,\\ 
Winterthurerstrasse 190, CH-8057 Z{\"u}rich, Switzerland\\
$^{2}$Department of Physics, ETH Zurich, Otto-Stern-Weg 1, CH-8093 Z{\"u}rich, Switzerland\\
$^{3}$Institute of Space Sciences (ICE, CSIC) \& Institut d'Estudis Espacials de Catalunya (IEEC) at Campus UAB,\\
Carrer de Can Magrans s/n 08193 Barcelona, Spain\\
$^{4}$Kavli Institute for Astronomy and Astrophysics at Peking University, 100871 Beijing, China\\
$^{5}$Institute of Applied Mathematics, Academy of Mathematics and Systems Science, Chinese Academy of Sciences, 100190 Beijing, China\\
$^{6}$Zentrum f{\"u}r Astronomie und Astrophysik, TU Berlin, Hardenbergstra{\ss}e 36, 10623 Berlin, Germany}

\date{Accepted 2020 May 04. Received 2020 April 14; in original form 2019 November 14}

\pubyear{2020}

\begin{document}

\label{firstpage}

\pagerange{\pageref{firstpage}--\pageref{lastpage}}

\maketitle


\begin{abstract}
We present a revised version of Peters' (1964) time-scale for the gravitational-wave (GW) induced decay of two point masses. The new formula includes the effects of the first-order post-Newtonian perturbation and additionally provides a simple fit to account for the Newtonian self-consistent evolution of the eccentricity. The revised time-scale is found by multiplying Peters' estimate by two factors, $R(e_0)= 8^{1-\sqrt{1-e_0}}$ and $Q_{\rm f}(p_0) = \exp \left(2.5 (r_{\rm S}/p_0) \right)$, where $e_0$ and $p_0$ are the initial eccentricity and periapsis, respectively, and $r_{\rm S}$ the Schwarzschild radius of the system. Their use can correct errors of a factor of 1--10 that arise from using the original Peters' formula. We apply the revised time-scales to a set of typical sources for existing ground-based laser interferometers and for the future Laser Interferometer Space Antenna (LISA), at the onset of their GW driven decay. We argue that our more accurate model for the orbital evolution will affect current event- and detection-rate estimates for mergers of compact object binaries, with stronger deviations for eccentric LISA sources, such as extreme and intermediate mass-ratio inspirals. We propose the correction factors $R$ and $Q_{\rm f}$ as a simple prescription to quantify decay time-scales more accurately in future population synthesis models. We also suggest that the corrected time-scale may be used as a computationally efficient alternative to numerical integration in other applications that include the modelling of radiation reaction for eccentric sources.
\end{abstract}

\begin{keywords}
Black hole physics -- Gravitational waves -- Methods: analytical
\end{keywords}


\section{Introduction}\label{sec:introduction}

The era of gravitational-wave (GW) astronomy started with the detection of merging stellar-mass black holes (BHs) and inspiralling neutron stars by the ground-based Laser Interferometer GW Observatory (LIGO) and Virgo detectors (\citealt{FirstGW,AbbottEtAl2017,GWTC-1}). These two detectors, along with the upcoming Laser Interferometer Space Antenna (LISA; \citealt{Amaro-SeoaneEtAl2017,Barack_et_al_2019}), not only allow physicists to observe the Universe, but offer an incredible tool to actively test the current theory of gravity: general relativity (GR). Making a prediction for a gravitational signal requires precise modelling of the physics of the source. Since the first successful numerical integration of the evolution of binary BH spacetime \citep[][]{Pretorius2005}, we have been in the position of simulating relativistic mergers in GR. The required integration is, however, extremely demanding for current computational resources, which makes covering a substantial part of the source's parameter space by means of fully relativistic simulations impractical. Moreover, different types of GW sources interact very differently with their astrophysical surroundings \citep[see, e.g.][]{Barausse2014}. These interactions must be modelled in order to produce realistic event- and detection-rate estimates. Fortunately, there have been major developments in the field of approximations to GR, which allow to model both the sources and their surroundings more efficiently than a fully relativistic simulation. Schemes such as the post-Newtonian (PN) expansion \citep[see, e.g.][]{Blanchet2014} or the effective one body approach \citep[see, e.g.][]{onelove} are used to find approximate solutions of Einstein's field equations and the associated equations of motion for the compact-object binary. They successfully describe the relevant features of fully relativistic orbits, such as perihelion precession and radiation of GWs, and can be used to cover a larger amount of parameter space at the loss of some precision. The latter process is especially relevant, as it determines how relativistic orbits shrink as they lose energy due to gravitational radiation.

The first successful quantitative description of the evolution of a binary's orbit by means of GW induced decay is due  to \cite{Peters_1964}, who used the quadrupole formula to calculate how the semimajor axis and the eccentricity of a Keplerian ellipse evolve in time. By taking these evolution equations, it is possible to find an approximate analytical estimate of the time-scale to the eventual binary's coalescence, if one does not take into account the self-consistent evolution of the eccentricity. This expression, commonly known as ``Peters' formula'' or ``Peters' time-scale'', is  exact for circular orbits and progressively less accurate for more eccentric ones. When a direct integration of the evolution equations is too computationally expensive, Peters' time-scale shines for its simplicity and predictive power, and is used in a wide range of applications to model the GW induced decay of various kinds of binary orbits \citep[see, e.g.][and many others]{farris,vanla,pau1,elisa}. At the present time, however, GW detectors such as LIGO/Virgo (and, in the future, LISA) are potentially able to probe eccentric orbits, for which Peters' formula is less accurate. Many GW sources are expected to radiate while retaining medium ($e \gtrsim 0.5$) to very high ($e\gtrsim0.9$) eccentricities \citep[e.g.][]{Amaro-Seoane2007,Antonini2016,Bonetti2016,Bonetti_et_al_2018,Khan_et_al_2018,Bonetti2019,Giacobbo2019,Giacobbo2019b}. Furthermore, the Newtonian analysis carried out by \citet{Peters_1964} is expected to fail for sources that will complete many GW cycles in strong-gravity regimes. In this paper, we address these issues and revise Peters' formula to make it more adequate to tackle these more extreme types of sources.

In Section~\ref{sec:methods}, we present the theoretical background of the problem, focussing on the PN series and the parametrization of relativistic orbits. In Section~\ref{sec:firstorder}, we derive and discuss an analytical PN correction to Peters' time-scale. We then address the problem of the self-consistent evolution of the eccentricity, as well as calibrate the analytical correction with more complete numerical calculations. In Section~\ref{sec:conclusions}, we discuss our findings and give a few examples for the most interesting sources of GWs, i.e. binaries of supermassive BHs (SMBHs), binaries of stellar-mass BHs, and extreme and intermediate mass-ratio inspirals (EMRIs and IMRIs). We speculate on the effects of the correction on event-rate estimates for GW detectors such as LIGO-Virgo and LISA.

\section{Theoretical background}\label{sec:methods}

In order to analytically describe the motion of two bodies in GR, it is necessary to approximate Einstein's field equations. An important and powerful approximation scheme is the Arnowitt--Deser--Misner formalism (hereafter ADM; \citealt{ADM_original}; see also \citealt{ADM_rep}). It offers a way to describe the motion of a binary system through the familiar concept of a Hamiltonian and Hamilton's equations for the separation and momentum vectors of the bodies. The ADM Hamiltonian, $H_{\rm ADM}$, is organised into a PN series. Here the PN order is denoted with the index $H_i$ and $H_{\rm N}$ denotes the Newtonian contribution:

\begin{align}
    H_{\rm ADM} = H_{\rm N} + \frac{1}{c^2}H_{1} + \frac{1}{c^4}H_{2} + \frac{1}{c^5}H_{2.5}(t) + \frac{1}{c^6}H_{3} + ..., \label{eq:ADM} \\ \nonumber
\end{align}

\noindent where $c$ is the speed of light in vacuum. Different gravitational effects (such as precession, spin-orbit coupling, etc.; see \citealt{Schaefer_rev} and references therein) enter the picture at their respective PN orders. Whereas all orders depend on the canonical variables, the 2.5 and all subsequent half-integer orders also explicitly depend on time. This is because half-integer orders describe energy dissipation due to GWs. The 2.5-term of the ADM Hamiltonian is determined by the quadrupole moment tensor of a Newtonian ellipse $\mathbf{N}$:

\begin{align}
    \frac{{\rm d}H_{2.5}(t)}{{\rm d}t} = \frac{G}{5 c^5}\frac{{\rm d}^3\mathbf{N}}{{\rm d}t^3}\frac{{\rm d}^3\mathbf{N}}{{\rm d}t^3}, \label{eq:25ham}\\ \nonumber
\end{align}

\noindent where $G$ is the gravitational constant. The explicit time dependence in the half-integer terms of the ADM Hamiltonian precludes the existence of a general analytic solution to Hamilton's equations. The integer-order part of the ADM Hamiltonian is often called the ``conservative part'' because it admits the definition of a conserved reduced (i.e. per unit mass) energy $E$ and, in certain cases (e.g. when the spin is zero), even a conserved reduced angular momentum $h$ \citep[][]{Schaefer_rev}.\footnote{For the remainder of the paper, energy and angular momentum have to be intended as reduced quantities, unless otherwise stated.} These conserved quantities have been used in \citet{1PN_par}, \citet{2PN_par}, and finally in \citet{QK_Parametrisation} to produce parametric solutions to the PN Hamilton's equations that are exact up to third order for binaries with no spin \citep[for another approach see, e.g.][]{yannick}. The solutions describe elliptical orbits undergoing periapsis precession; a simple way to describe them is to use the familiar concepts of semimajor axis, $a$, and eccentricity, $e$. However, some care is needed because the simple Newtonian relations between energy, angular momentum, and the orbital parameters are no longer valid in the PN model. Rather, the definitions of the orbital parameters are now  given as PN series:

\begin{align}
    a &= -\frac{G M}{2 E} + \frac{a_{\rm 1\,PN}(E,h)}{c^2} + \frac{a_{\rm 2\,PN}(E,h)}{c^4} + ..., \label{eq:frol}\\
    e^2 &= 1 + E h^2 + \frac{e^2_{\rm 1\,PN}(E,h)}{c^2}  + \frac{e^2_{\rm 2\,PN}(E,h)}{c^4} + ...,\label{eq:fral}\\ \nonumber
\end{align}

\noindent where $M = m_1 + m_2$ (with $m_1 \geq m_2$) is the combined mass of the two bodies. These orbital parameters (along with the periapsis precession) can be used to describe the third-order PN behaviour of the orbit in absence of the dissipative gravitational radiation term (2.5~PN). There is, however, a way to reintroduce dissipation effects in the description of the orbit. To understand it, we briefly return to the case of a Newtonian orbit.

The energy $E$ and angular momentum $h$ are conserved quantities of a Newtonian binary orbit. They completely determine its shape and size by means of the formulae

\begin{align}
    a &= -\frac{G M}{2 E}, \label{eq:aaa}\\
    e^2 &= 1 + 2h^2E. \label{eq:eee}\\ \nonumber
\end{align}

On the other hand, Einstein's quadrupole formula \citep[][]{GR} predicts that the (non-reduced) energy $\mathcal{E}$ and angular momentum $L$ will be radiated from a gravitational system in the form of GWs:

\begin{align}
    \frac{{\rm d}\mathcal{E}}{{\rm d}t}&=\frac{G}{5c^5}\frac{{\rm d}^3M_{ij}}{{\rm d}t^3}\frac{{\rm d}^3M_{ij}}{{\rm d}t^3}, \label{eq:quad}\\
    \frac{{\rm d}L}{{\rm d}t}&= \epsilon_{zab}\frac{{\rm d}^2M_{ac}}{{\rm d}t^2}\frac{{\rm d}^3M_{bc}}{{\rm d}t^3}.\\ \nonumber
\end{align}

\indent Here the matrix $\mathbf{M}$ contains the components of the mass quadrupole moment tensor of the binary for an angular momentum pointing in the $z$ direction. There is a clear inconsistency between the conserved quantities of a purely Newtonian description and the dissipation equations of the quadrupole formula. This inconsistency can be alleviated by noting that changes in $\mathcal{E}$ and $L$ occur on time-scales much longer than the typical orbital (radial) period. Indeed, the time-scale over which the (non-reduced) energy changes scales as ${\rm d}t \propto c^5 d\mathcal{E}$, whereas the typical orbital period is independent of $c$. This consideration is what allowed \citet{Peters_1964} to combine Keplerian celestial mechanics with the quadrupole formula to obtain the equations for the GW induced evolution of orbital parameters:

\begin{align}
    \frac{{\rm d}a}{{\rm d}t} &= - \frac{64}{5c^5}\frac{G^3M^3 q}{a^3(1 + q)^2} f(e), \label{eq:at}\\
    \frac{{\rm d}e}{{\rm d}t} &= -e\frac{304}{15c^5}\frac{G^3M^3 q}{a^4(1-e^2)^{5/2}(1 + q)^2}\left(1 + \frac{121}{304}e^2\right), \label{eq:et}\\ \nonumber
\end{align}

\noindent where

\begin{align}
    f(e)&= \left(1 + \frac{73}{24}e^2 +\frac{37}{96}e^4\right)(1-e^2)^{-7/2} \label{eq:enh}\\ \nonumber
\end{align}

\noindent and $q = m_2/m_1\leq 1$ is the mass ratio. These equations describe how a Newtonian orbit slowly changes shape if its energy and angular momentum are radiated away according to the quadrupole formula. By manipulating Equations~\eqref{eq:at} and \eqref{eq:et}, while also assuming that the eccentricity remains fixed at its initial value, it is possible to find an analytic expression for the time-scale, $t_{\rm P}$, over which the orbit decays. The resulting expression is known as ``Peters' time-scale'' in the literature. It scales with the fourth power of the initial Keplerian semimajor axis, $a_0$, weighted by the `eccentricity enhancement function' $f(e_0)$, $e_0$ being the initial Keplerian eccentricity:

\begin{align}
    t_{\rm P} = \frac{5c^5(1 + q)^2}{256 G^3 M^3q} \frac{a_0^4}{f(e_0)}. \label{eq:pet}\\ \nonumber
\end{align}

The assumption used in \citet{Peters_1964} is that, despite the radiation of energy, the binary is still moving along the orbit predicted by Newtonian mechanics. Note however, that the results of the quadrupole formula (Equation~\ref{eq:quad}) in the context of Peters' work and the 2.5~PN Hamiltonian term (Equation~\ref{eq:25ham}) in the context of the ADM formalism are identical. Indeed, there are no additional terms in the ADM Hamiltonian that describe energy radiation before the 3.5~PN order. This suggests that the quadrupole formula should be able to describe the orbital evolution not only of a Newtonian orbit, but also of a PN orbit up to third order.
 
To prove this statement, one can attempt to ``improve'' the results of the quadrupole formula. We start with a simple Newtonian orbit and insert it in Equation~\eqref{eq:quad}. We then obtain an evolution equation for the energy $E$ that scales with $c^{-5}$. As noted before, this is already equivalent to the 2.5~PN ADM Hamiltonian. From here on, there are only two possibilities to improve the accuracy of the description of energy radiation via GWs: the contribution of a 1~PN perturbed orbit to quadrupolar radiation and the contribution of a Newtonian orbit to octupolar radiation. To improve it even further, it is necessary to add more and more contributions \citep[see, e.g.][for more details on this topic]{thorne, radrec,Blanchet2014}. We show the process in a schematic form:

\begin{align}
    \frac{{\rm d}E}{{\rm d}t} &= \text{$2^2$pole[Newton] +  first tail contribution}  \nonumber \\ &+ \text{($2^2$pole[1~PN] + $2^3$pole[Newton])} \nonumber\\
    &+ \text{($2^2$pole[2~PN] + $2^3$pole[1~PN] + $2^4$pole[Newton])} \nonumber\\
    &+ \text{higher-order terms}, \\ \nonumber
\end{align}

\noindent where we denote the operation of calculating the $2^n$pole contribution of an orbit with the functions ``$2^n$pole[ ]''. The functions scale with $c^{-1 - 2n}$. We also denote the various perturbations to the orbital dynamics as ``Newton, 1~PN, 2~PN, ..., $n$~PN''. The perturbations scale as $c^{-2n}$. The first tail contribution describes the coupling between the mass quadrupole and the mass monopole and is of one and a half orders higher than the purely quadrupolar radiation. Note how any improvement to the results of the quadrupole formula will necessarily have at least a $c^{-7}$ factor. Indeed, any improvement to the quadrupole formula is of 3.5~PN order. The situation is very different if we consider the assumption of Keplerian orbits: perturbations already arise at the first PN order. Another way to state this is the following: the assumption of Keplerian orbits fails at a lower PN order than the assumption of quadrupolar energy radiation.

The idea of this paper is thus to {\it apply the quadrupole formula to the energy of a PN orbit} rather than a Keplerian orbit. The parametric solution to the conservative ADM Hamiltonian equations explicitly depends on some kind of conserved energy and angular momentum, analogous but not necessarily equal to their Newtonian counterparts. By taking these quantities and using them as initial conditions for the energy dissipation equations, we can improve the results derived in \citet{Peters_1964} without needing any higher-order energy flux term. We are aware that truncating the ADM Hamiltonian at 2.5~PN order is a simplification, as tail effects and 3.5~PN terms in the energy flux are considered to be of  ``relative first order'' with respect to the quadrupole formula. If included, they would contribute to the secular evolution of the 1~PN semimajor axis and eccentricity. However, the truncation at 2.5~PN is justified in the formal sense of a Taylor expansion of the ADM Hamiltonian. From this point of view, the higher-order contributions can be interpreted as being small corrections to the overall behaviour of the PN orbital parameters, which is dictated by the quadrupole formula alone (up to order 3~PN included). At the cost of some precision, this simplification (among others) will be absolutely necessary to find an analytic correction to Peters' formula, valid for a large region of parameter space.

In Sections~\ref{sec:fix2} and \ref{sec:sig}, we compute how much these considerations alter the original prediction of \citet{Peters_1964} for the duration of a binary decay and discuss the analytic results. In Section~\ref{subsec:consistentPN}, we perform a numerical analysis of the effect of the 3.5 PN fluxes and re-calibrate the analytical formula to account for them.

\section{Derivation of the Corrections}\label{sec:firstorder}

\subsection{Orbits with identical initial shape}\label{sec:fix2}

The goal of this section is to derive an explicit analytic correction to Peters' time-scale in the PN framework. In order for the correction to be  analytic, we have to restrict our calculations to the lowest significant order, effectively combining  the $c^{-2}$ conservative information with the $c^{-5}$ non-conservative information.

In their original Newtonian form, Equations~\eqref{eq:at}--\eqref{eq:et} simultaneously describe the change of the geometry and that of the energy state of a binary system. The introduction of PN effects breaks this symmetry, and it becomes necessary to carefully compare appropriate initial conditions. In the typically adopted expression for Peters' time-scale, one usually computes the orbital decay time by plugging in an initial (Keplerian) semimajor axis and eccentricity. Hence, we will express our correction in the same terms.

We start by recalling the explicit first-order expressions for the semimajor axis, $a_{1}$,  and eccentricity, $e_{1}$, of the perturbed orbit (that describe a more realistic orbital shape, i.e. closer to the relativistic result), in terms of the PN conserved energy $E_1$ and angular momentum $h_1$ \citep{QK_Parametrisation}:

\begin{align}
   a_1 &= -\frac{G M}{2 E_1}-\frac{G M \left(7 q ^2+13 q
   +7\right)}{4 c^2 (1+q)^2}, \label{eq:2}\\
   e_{1}^2 &= 1+ 2 E_{1} h_{1}^2 \nonumber \\ 
   &+ \frac{E_1\left(5 E_1 h_1^2 \left(3 q^2+5 q +3\right)+2 \left(6
   q ^2+11 q +6\right)\right)}{c^2 (1+q )^2}.\label{eq:2B}\\ \nonumber
\end{align}

We want to investigate the difference in the decay time-scale when we compare Newtonian and PN orbits with the same initial ``shape''. This amounts to setting the perturbed orbital parameters $a_1$ and $e_1$ equal to the Newtonian ones, $a_{\rm N}$ and $e_{\rm N}$. In the case of a first-order PN analysis, the equations that one obtains are

\begin{align}
    a_1 &\overset{!}{=} a_{\rm N}, \label{eq:a} \\
    e_1^2 &\overset{!}{=} e_{\rm N}^2. \label{eq:b}\\ \nonumber
\end{align}

By rearranging these expressions, we obtain a set of conditions for the PN energy, $E_1$, and the PN angular momentum, $h_1$. We can interpret the resulting conditions as the values that $E_1$ and $h_1$ must have in order for the perturbed orbit to be identical to a Newtonian orbit with semimajor axis $a_{\rm N}$ and eccentricity $e_{\rm N}$. The conditions read

\begin{align}
    E_1&=-\frac{G M}{2 a_{\rm N}}+\frac{G^2 M^2 \left(7 q ^2+13 q +7\right)}{8 a_{\rm N}^2 c^2
   (q +1)^2}+O\left(\frac{1}{c^3}\right), \label{eq:c} \\
   h_1^2&= \frac{a_{\rm N}(1- e_{\rm N}^2)}{G M}  \nonumber \\ &+ \frac{4(q +1)^2+e_{\rm N}^2 \left(2 q ^2+3 q +2\right)}{c^2(q +1)^2}+O\left(\frac{1}{c^3}\right). \label{eq:d}\\ \nonumber
\end{align}

Equations~\eqref{eq:c} and \eqref{eq:d} show that the perturbed orbit has a slightly larger energy and squared angular momentum than the non-perturbed one. In other words, it takes more energy and more angular momentum to sustain a certain shape in a PN gravity field rather than in a Newtonian gravity field. Since excess energy and momentum take time to be radiated away, we come to a qualitative result: \textit{PN orbits decay more slowly than what Peters' formula predicts, when comparing binaries with identical initial shape (i.e. semimajor axis and eccentricity).}

This result might seem counter to the common expectation that PN effects reduce the time-scale of GW induced decay. This apparent discrepancy is due to the fact that a more common way to confront Newtonian and PN results is to compare orbits with identical initial energy and angular momentum \citep[e.g.][]{Blanchet2014}, and not identical initial semimajor axis and eccentricity. As soon as one includes PN contributions, comparing orbits with the same initial shape or energy and angular momentum is not equivalent any longer. In Appendix~\ref{sec:fix1}, we show the same analysis for the latter comparison, which indeed produces different results. To quantify the time-scale for the decay, it is useful to define two new parameters, an effective semimajor axis, $a_{\rm eff}$, and an effective eccentricity, $e_{\rm eff}$:

\begin{align}
    a_{\rm eff} &=- \frac{G M}{2 E_1},\\
    e_{\rm eff}^2 &= 1 + 2 h_1^2 E_1.\label{eq:effe_def}\\ \nonumber
\end{align}

These ``effective'' parameters are used to construct a fictitious Newtonian orbit that contains all the information regarding the perturbed orbital parameters. Their explicit formulae are

\begin{align}
    a_{\rm eff} &= a_{\rm N}+ \frac{G M \left(7 q ^2+13 q +7\right)}{4 c^2 (q +1)^2}, \label{eq:effa} \\
    e_{\rm eff}^2  &=e_{\rm N}^2 \label{eq:effe} \\ \nonumber &- \frac{G M \left(5 e_{\rm N}^2 \left(3 q ^2+5 q +3 \right)+9 q ^2+19 q +9\right)}{4 c^2
   a_{\rm N} (q +1)^2}. \\ \nonumber
\end{align}

Note that these effective quantities do not carry information on the actual shape of the orbit; we introduce them as convenient variables for the analysis. Equations~\eqref{eq:effe} and \eqref{eq:effa} imply that: (i) the slight increase of the perturbed energy $E_1$ with respect to the Newtonian energy causes the effective semimajor axis to be larger than the actual one; (ii) the slight increase of the perturbed squared angular momentum $h_1^2$ causes the squared effective eccentricity to decrease; and (iii) the squared effective eccentricity may in principle become negative (indeed, it is always negative if the physical orbit is circular). 

\begin{figure}
    \centering
    \includegraphics[scale=0.67]{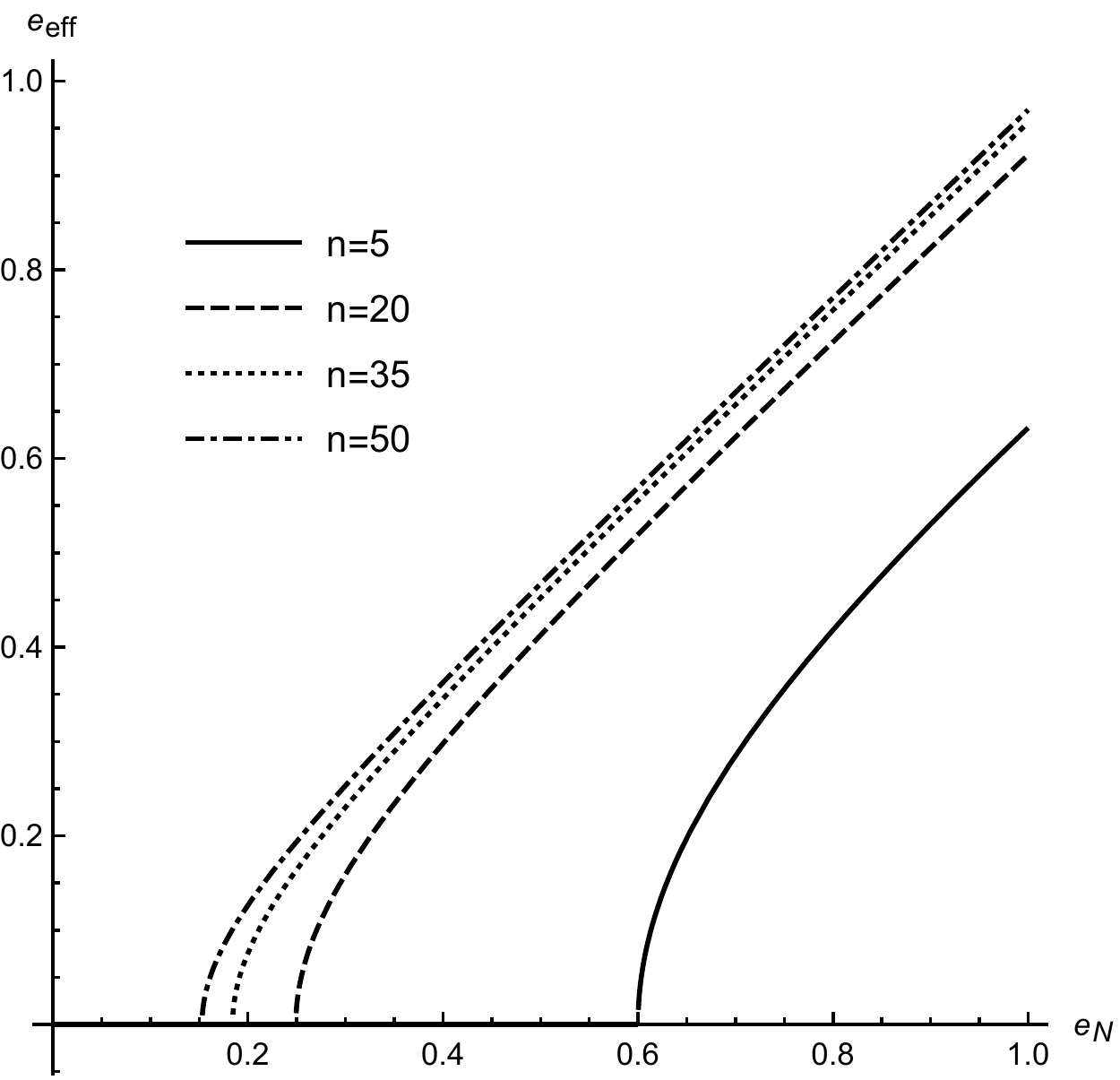}
    \caption{We plot the effective eccentricity (Equation~\ref{eq:effe}) against the Newtonian eccentricity for orbits with different initial sizes ($a_{\rm N} = n r_{\rm S}$). The effective eccentricity is always slightly smaller than its Newtonian counterpart. The significance of this difference is determined by the orbit's distance to the Schwarzschild radius of the system: it decreases for larger orbits, as expected for any PN perturbation. In very strong gravity regimes, there is no way to describe the energy and angular momentum of the orbit as an effective Newtonian ellipse. In the plot, we see this effect when the effective eccentricity becomes zero.
    }
    \label{fig:effe}
\end{figure}

Summing up, the effective ellipse is larger and more circular than the actual (PN) orbit because it is constructed using the larger PN energy and angular momentum. However, we incur into a problem if the physical orbit is already very circular. The crucial point is that PN orbits may have more angular momentum than what the classical definition of eccentricity and semimajor axis allow for a given energy. In other words, if the physical orbital eccentricity is smaller than a critical value $e_{\rm c}$, no Newtonian orbit exists with corresponding energy and angular momentum, as implied by the possible appearance of a negative ``squared eccentricity''. In  Figure~\ref{fig:effe}, we show the effective versus the Newtonian eccentricity for orbits with different initial sizes. There exists, however, a simple solution to this apparent inconsistency, which extends the validity of the effective orbit construction to all possible perturbed orbits. Note that, if negative, the absolute value of the squared effective eccentricity is always at least smaller than a 1~PN term:

\begin{align}
   \lvert e_{\rm eff}^2 \rvert &= \left| e_{\rm N}^2 - \frac{G M \left(9 + 15 e_{\rm N}^2 \right) }{4 c^2 a_{\rm N}} +O\left(q\right) \right| < \left| \frac{6 G M}{ c^2 a_{\rm N}} \right|. \label{eq:effe2} \\ \nonumber
\end{align}

Therefore, if we allow for a small adjustment (smaller than a 1~PN term) to the squared angular momentum of the effective ellipse, we can neglect the cases where the squared effective eccentricity would be negative and instead treat the effective ellipse as having zero eccentricity.

The time-scale for the orbit's decay can now be computed by integrating Equation~\eqref{eq:at} from the value $a_{\rm eff}$ to a value representing the coalescence of the binary, which we simply take to be the effective Schwarzschild radius of the two-body system (hereafter, simply Schwarzschild radius), $r_{\rm S} = 2GMc^{-2}$.

In order to solve the decay Equation~\eqref{eq:at} analytically, we have to make the simplifying assumption of small initial eccentricity ($e_{\rm N}^2 \approx 0$). For a given initial semimajor axis $a_0$, the time evolution of a circular orbit reads:

\begin{align}
    a_{\rm N}(t)= \sqrt{2} \left( a_0^4-\frac{256 G^3 M^3 q  t}{5 c^5 (q +1)^2}\right)^{\frac{1}{4}}. \label{eq:3}\\ \nonumber
\end{align}

With this formula, we can compute the corrected time-scale, $\tau_{\rm c}$, for the GW decay of a perturbed orbit by replacing the initial semimajor axis $a_0$ with the effective semimajor axis, and subsequently solving Equation~\eqref{eq:3} for $t$. The resulting expression would only be valid for circular orbits. Nonetheless, there is a simple way to reintroduce large initial eccentricities. Note that, since the time-scale scales with the fourth power of $a_0$, most of the decay time is spent in the neighbourhood of the initial conditions. Therefore, we can adjust the corrected time-scale by dividing it by the eccentricity enhancement function $f(e)$ evaluated at the initial effective eccentricity:

\begin{align}
    \tau_{\rm c} \to \frac{\tau_{\rm c}}{f(e_{\text{eff}})}.\\ \nonumber
\end{align}

This manipulation is essentially the same one that is used in \cite{Peters_1964} to yield what we refer to in this paper as Peters' formula (Equation \ref{eq:pet}).

The results of the calculations are most readable when expressed in terms of multiples $a_0 = n r_{\rm S}$ of the Schwarzschild radius:

\begin{align}
 \tau_{\rm c}&=\frac{5 (q +1)^2  \left(\left(n+\frac{7 q ^2+13 q +7}{8 (q
   +1)^2}\right)^4-\frac{1}{4}\right)r_{\rm S}}{32 c q f(e_{\rm eff}) }. \label{eq:tauc}\\ \nonumber
\end{align}

We report the ratio $\nu_{\rm g} = \tau_{\rm c}/t_{\rm P}$ and difference $\delta_{\rm g} = \tau_{\rm c}-t_{\rm P}$ between time-scales in the two limiting cases of $q \to 0$,

\begin{align}
    \nu_{\rm g} &=\frac{4 \left(n+\frac{7}{8}\right)^4-1}{4 n^4-1} \frac{f(e_0)}{f(e_{\rm eff})} \ge 1, \label{eq:nu_g} \\
    \delta_{\rm g} &=\frac{5 r_{\rm S} }{32 c q }\left( \frac{
   \left(n+\frac{7}{8}\right)^4}{f(e_{\rm eff})} - \frac{ n^4}{f(e_{0})} \right) \ge 0, \label{eq:delta_g}\\ \nonumber
\end{align}

\noindent and $q \to 1$,

\begin{align}
      \nu_{\rm g} &=\frac{4 \left(n+\frac{27}{32}\right)^4-1}{4 n^4-1} \frac{f(e_0)}{f(e_{\rm eff})} \ge 1, \label{eq:nu_glargeq} \\
    \delta_{\rm g} &=\frac{5 r_{\rm S} }{32 c }\left( \frac{4 n^4-\frac{12117361}{1048576}}{f(e_{\rm eff})} - \frac{4 n^4-1}{f(e_{0})} \right) \ge 0, \label{eq:delta_glargeq}\\ \nonumber
\end{align}

\noindent computed for the same initial orbital shapes.

\subsection{Simplifying the analytic result}\label{sec:sig}

In the previous section, we defined two quantities ($\nu_{\rm g}$ and $\delta_{\rm g}$) that describe the deviations from Peters' formula when taking PN perturbations to the orbit into account and comparing orbits with identical initial shape. Now, let us take a closer look at the ratio of the corrected to the standard Peters' time-scale.

\begin{figure}
    \centering
    \includegraphics[scale=0.67]{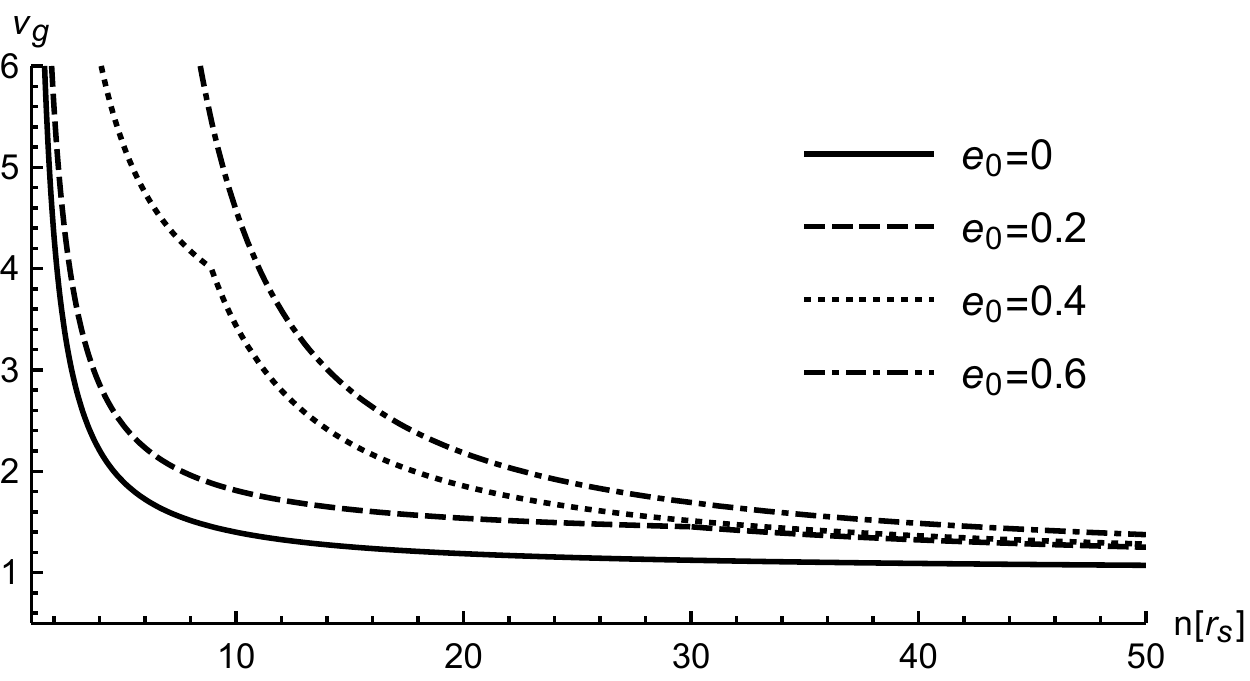}
    \includegraphics[scale=0.67]{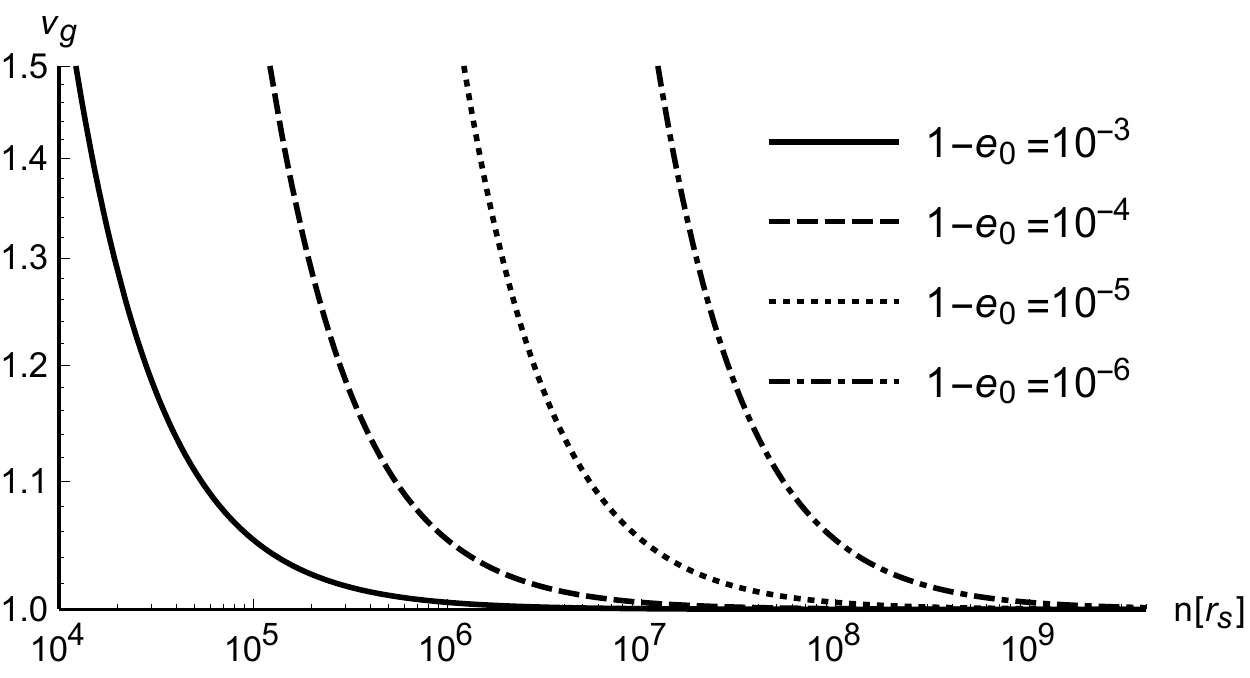}
    \caption{We show the behaviour of the PN-correction to Peters' time-scale ratio $\nu_{\rm g}$ plotted against the initial semimajor axis measured in Schwarzschild radii ($a_0 = n r_{\rm S}$). Since the ratio $\nu_{\rm g}$ depends on the orbital eccentricity, we plot it for different values of the initial eccentricity $e_0$. $\nu_{\rm g}$ approaches unity for $n \to \infty$, as expected, but can still be significantly different from one for very large $n$ if the initial orbit is very eccentric, suggesting that the magnitude of this correction is sensitive to the periapsis of the system.}
    \label{fig:nu}
\end{figure}

The ratio $\nu_{\rm g}$ approaches unity for $n$ approaching infinity. This is expected, because any PN perturbation to the binary orbit has to vanish as we increase the relative distance between the two bodies. However, a quick glance at Figure~\ref{fig:nu} shows that, as $n$ increases, the ratio $\nu_{\rm g}$ remains quite large for a wide range of $n$. More specifically, the behaviour of $\nu_{\rm g}$ (which is also a function of the initial orbital eccentricity) strongly depends on the value of the initial orbital periapsis, $p_0 = a_0(1-e_0)$: the smaller the periapsis, the more enhanced the deviation from Peters' time-scale.

The explicit analytical formula of $\nu_{\rm g}$ is still rather complex. It is possible to find a much simpler formula that can approximate it very well by taking the limit of very large $n$ (large orbits) and $e_0$ approaching unity (very eccentric orbits):

\begin{align}
 \nu_{\rm g} \to 1+ \frac{7 \left(6 q^2+11 q+6\right)}{8 \left(1-e_0\right) n (q+1)^2}.  \label{eq:approx_nug}\\ \nonumber
\end{align}

\begin{figure}
    \centering
    \includegraphics[scale=0.67]{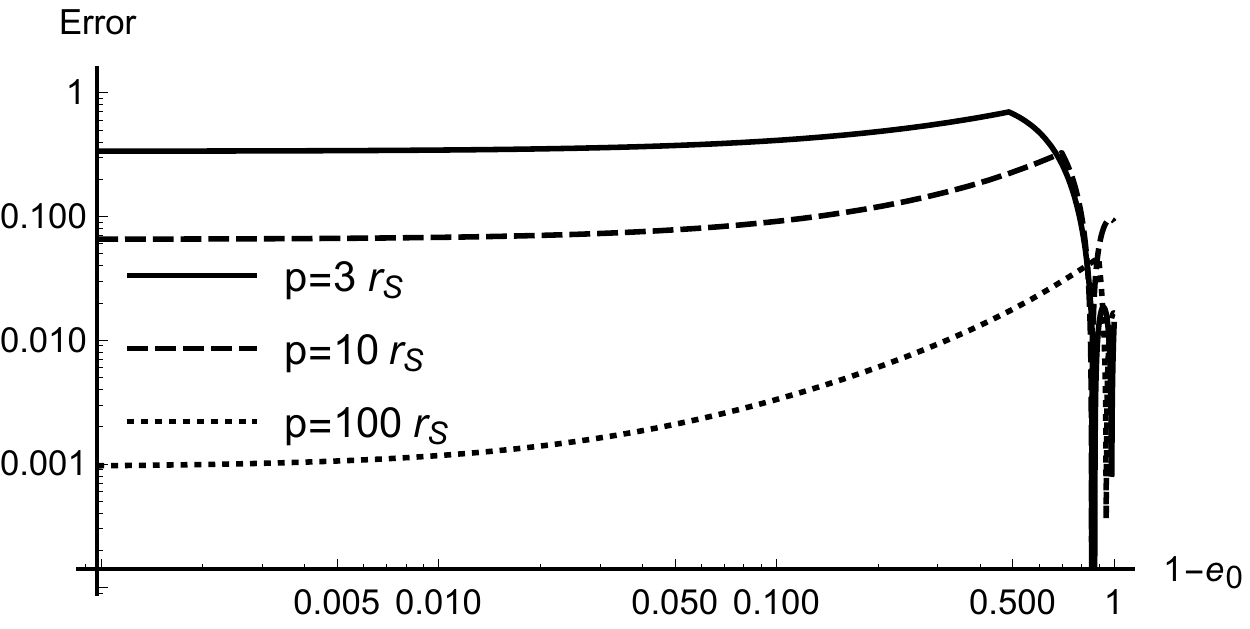}
    \caption{We show the relative error of the simplified ratio $Q$ (Equation~\ref{eq:Q}) to the exact first-order correction $\nu_{\rm g}$ (Equation~\ref{eq:tauc} divided by $t_{\rm P}$), for a wide range of initial conditions. For very eccentric orbits ($1 - e_0 < 0.1$), the error is of the magnitude of a PN term $r_{\rm S}/ p_0$, where $p_0$ is the periapsis. This implies, that the absolute error that arises by using this approximation instead of $\nu_{\rm g}$ is as large as a second-order PN term and can be neglected. The relative error increases for orbits with intermediate eccentricity ($1 - e_0 \approx 0.5$) but dramatically falls for almost circular orbits ($1 - e_0 \approx 0.8$). It also shows a complicated behaviour for very low eccentricities but always remains small.
    }
    \label{fig:error}
\end{figure}

It turns out that the above approximation actually covers most of the phase space without producing a significant loss of precision. This aspect can be appreciated by looking at Figure~\ref{fig:error}, which shows the discrepancy between the approximated $\nu_{\rm g}$ (Equation~\ref{eq:approx_nug}) and the exact 1~PN  value (Equation~\ref{eq:nu_g}). In short, the absolute error of the approximation with respect to the full 1 PN result is of the order of a 2 PN term, and can therefore be neglected.

We can rewrite Equation~\eqref{eq:approx_nug} as

\begin{align}
    \nu_{\rm g} = 1 +\frac{7 \left(6 q^2+11 q+6\right) r_{\rm S}}{8 \left(1-e_0\right) a_0 (q+1)^2} = 1+K(q) \frac{r_{\rm S}}{a_0(1-e_0)} \ge 1. \label{eq:nug}\\ \nonumber
\end{align}

The PN correction factor can be easily interpreted: it shows that the time-scale of gravitational decay is affected in a significant way by PN perturbations if the orbital periapsis $p$ comes close to the Schwarzschild radius of the system. The magnitude of the correction scales as $K \times (r_{\rm S} /p_0)$, with $K(q)$ being a monotonic function of the mass ratio $q$ that decreases from 5.25 (or 21/4) to approximately 5.03 (or 161/32) as $q$ ranges from 0 to 1. The variation is very weak because the first-order perturbations to the semimajor axis and eccentricity only weakly depend on the mass ratio. We therefore name the approximation value for the above ratio the ``PN correction factor'' $Q$:

\begin{align}
    Q(p_0) = 1+5 \frac{r_{\rm S}}{a_0(1-e_0)} \ge 1. \label{eq:Q}\\ \nonumber
\end{align}

By using this simple factor, we can restate the qualitative result of Section~\ref{sec:fix2} in a more precise form, valid when comparing orbits with identical initial semimajor axis and eccentricity, within the assumptions of this section: \textit{Peters' time-scale is too short by a factor of $Q$ that is relevant whenever the periapsis $p_0$ of the binary comes close to the Schwarzschild radius $r_{\rm S}$. The factor is given by the formula $Q=1 + 5 r_{\rm S} p_0^{-1}$.}

\subsection{Inclusion of the secular evolution of the eccentricity}\label{subsec:consistentecc}

In the previous sections, we neglected the self-consistent evolution of the eccentricity, in order to keep the method analytical, and used the common practice of dividing the decay time for circular orbits by the eccentricity enhancement function $f(e_0)$. This approximation is known to break down for orbits that start with very high eccentricities and can produce very large errors (of the order of 0, $\sim$400, and $\sim$700 per cent for initial eccentricities of 0, 0.9, and 0.999, respectively) in the evaluation of the decay time-scale. It is mostly used as a lower bound estimate for the self-consistently evolved case. Even in a purely Newtonian framework, the only way to achieve more accurate results is to integrate Equations~\eqref{eq:at} and \eqref{eq:et} directly and obtain a numerical value for the decay time-scale. While the numerical integration is straightforward and fast on a case by case basis, it is much less convenient in applications that require a quick and simple estimate for the decay time-scale of many orbits.

\begin{figure}
    \centering
    \includegraphics[scale=0.75]{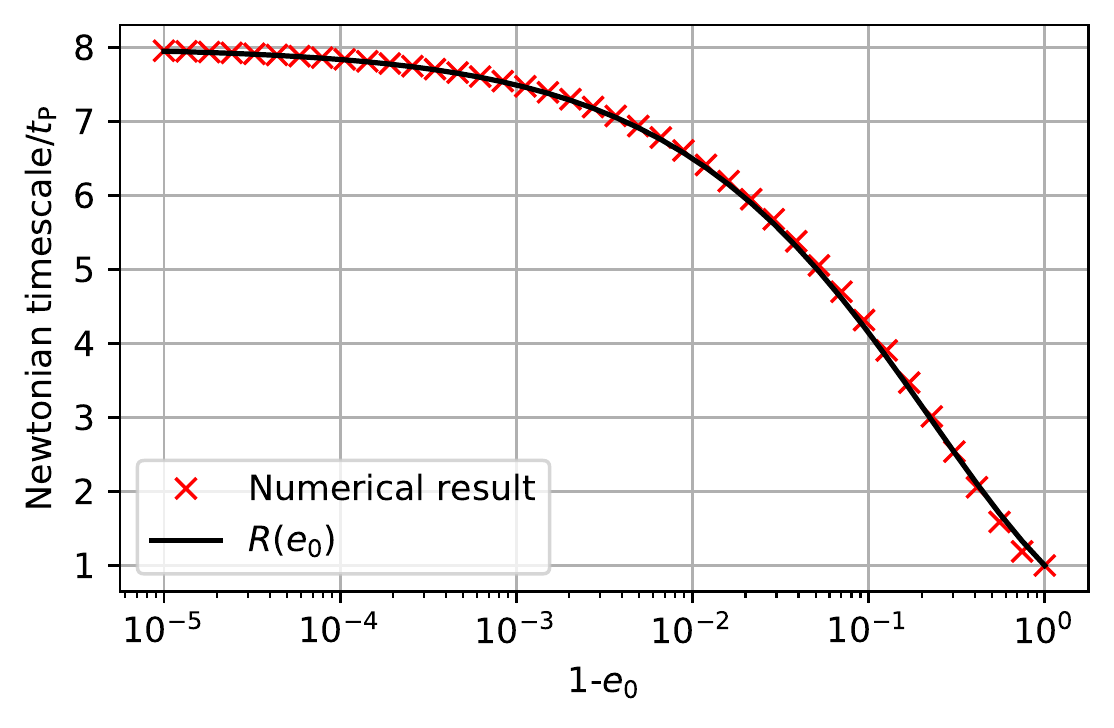}
    \includegraphics[scale=0.75]{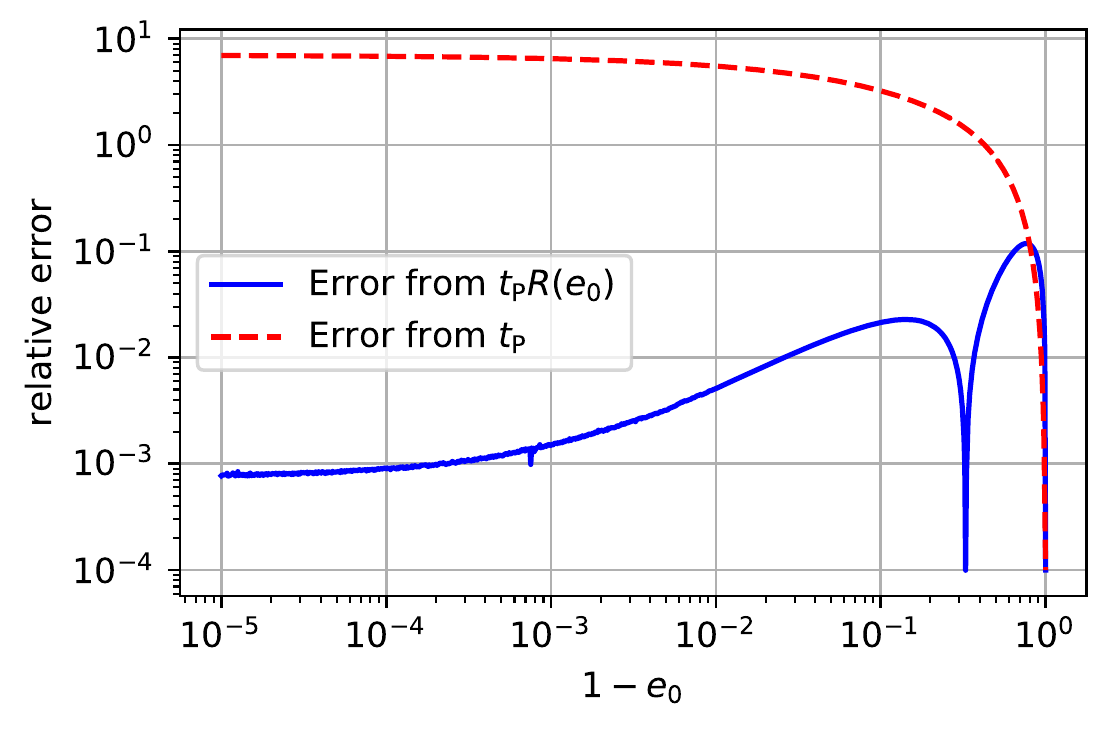}
    \caption{The top panel shows the ratio of the decay time-scale obtained by integrating the evolution Equations~$\eqref{eq:at}$--\eqref{eq:et} to Peters' formula (red crosses). The black solid line shows the fitting function $R(e_0)=8^{1-\sqrt{1-e_0}}$. The bottom panel shows the relative error from the numerical result that one incurs into when using Peters' formula (red dashed line) and when using the corrected formula $R(e_0)t_{\rm P}$ (blue solid line). The corrected time-scale reproduces the true result within a few per cent accuracy for a large region of parameter space, whereas Peters' formula dramatically fails (relative error of order $\sim$100 per cent) as soon as the eccentricity is larger than $\sim$0.5.}
    \label{fig:refit}
\end{figure}

In this section, we present a simple correction to Peters' formula that extends its validity to highly eccentric orbits in a Newtonian-gravity regime. To find this correction, we have performed numerous fully consistent integrations of Equations~\eqref{eq:at} and \eqref{eq:et} to find an accurate decay time-scale. We define the correction factor $R$ as the ratio between such accurate Newtonian time-scales and the lower bound given by Peters' formula (i.e. Equation~\ref{eq:pet}). It turns out that expressing the deviations in this manner cancels out the effects of most of the parameters: there is no appreciable variation in the value of the ratio $R$ when changing the mass ratio, the total mass, or the initial semimajor axis.

A plot of $R(e_0)$ and our proposal for a convenient fitting function are shown in the top panel of Figure \ref{fig:refit}. We found a surprisingly performing fitting function of the form $R(e_0)= A ^{ (1-(1-e_0)^{m})}$. The best-fitting parameters, when using 50 equally spaced points in $\log_{10}(e_0)$ between $e_0 = 0$ and 0.9999, are $A = 8.023$ and $m = 0.502$. The variation in the best-fitting values when choosing different semimajor axes, total masses, or mass ratios is limited to the third decimal place. By using round numbers, and without any great loss of accuracy, we adopt

\begin{align}
    R(e_0) = 8^{1-\sqrt{1-e_0}}. \label{eq:R} \\ \nonumber
\end{align}

The bottom panel of Figure~\ref{fig:refit} shows the error that arises from using Equation~\eqref{eq:R} with respect to the numerical results. It is clear that the proposed fit is an accurate estimation of the numerical result, especially for eccentric orbits ($e_0 \gtrsim 0.5$), where the relative error falls below a few per cent. For the sake of clarity, we stress that we do not claim that this is the best possible fit for the given data as a function of eccentricity. Indeed, for more circular orbits, we incur into some errors of fractional order ($\lsim$10 per cent), comparable to the performance of Peters' formula. However, we can claim that the fitted ratio $R(e_0)$ \textit{is sufficiently accurate and simple to extend the validity of Peters' formula to a wide range of initial eccentricities.} The ratio $R(e_0)$ can be thought as an extension of the eccentricity enhancement function. We can thus use the original Peters' formula (Equation~\ref{eq:pet}), after replacing $f(e_0)$ with the `effective eccentricity enhancement function' $f(e_0)/R(e_0)$. The corrected Newtonian formula can be manipulated just as easily as Peters' formula, can correct errors as large as a factor of eight, and it is equivalent to the integration of the two coupled evolution Equations~\eqref{eq:at} and \eqref{eq:et}.

\subsection{Calibration of the PN correction with 3.5 PN fluxes}\label{subsec:consistentPN}

\begin{figure}
    \centering
    \includegraphics[scale=0.75]{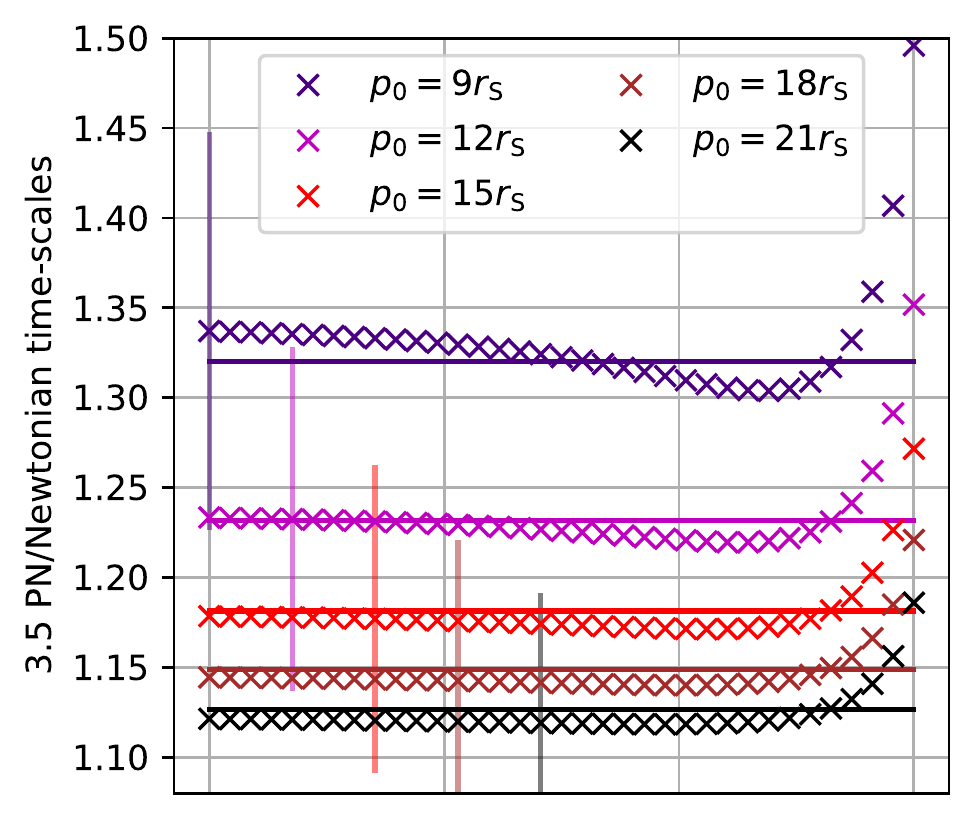}
    \includegraphics[scale=0.75]{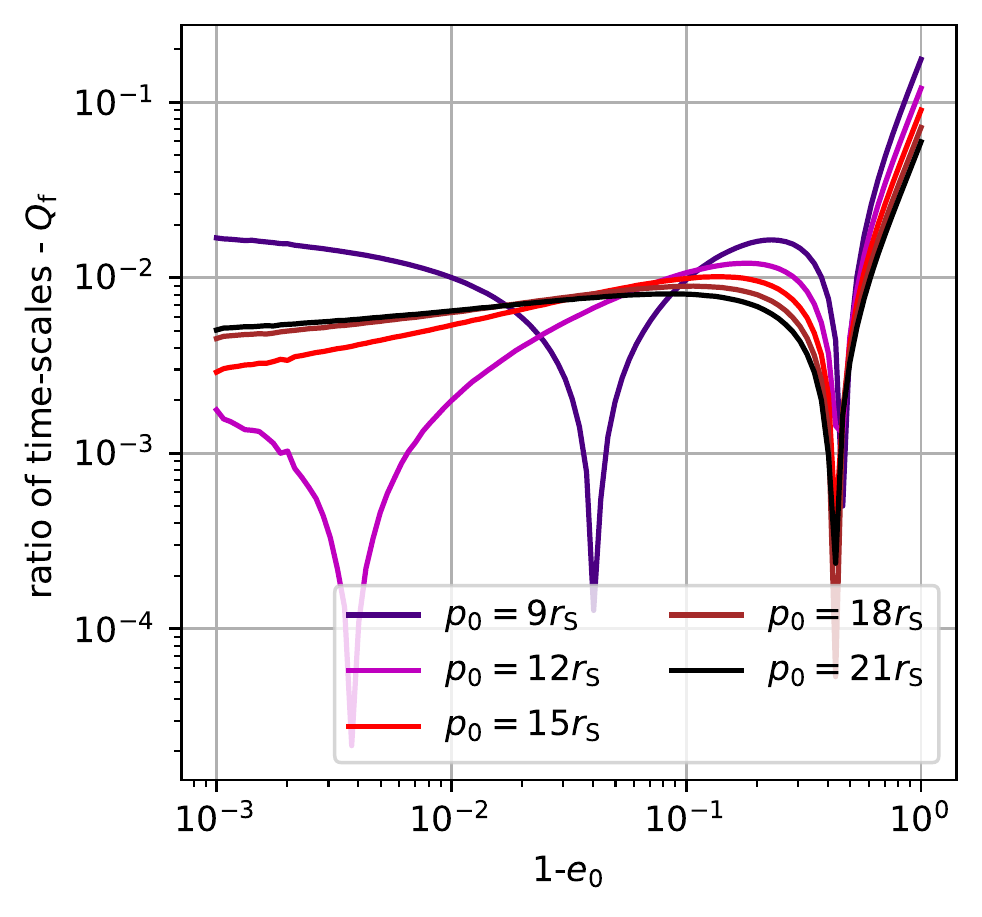}
    \caption{In the top panel, we divide the result from the full 3.5 PN evolution equations by the result from the full 2.5 PN evolution equations (crosses), as a function of initial eccentricity, for different initial periapses. We also show the proposed fits (solid horizontal lines) and a representative uncertainty due to the PN tail perturbations (vertical error bars; see text for the derivation). In the bottom panel, we plot the relative error that one would incur into by using the fit instead of the true 1~PN result. The errors are well below the typical uncertainty due to the PN tail contributions.}
    \label{fig:pnfit}
\end{figure}

In the derivation of the analytical PN correction factor $Q$, we treated the 3.5~PN fluxes (which are of first relative order to the Newtonian fluxes) as a negligible correction in the Taylor expansion of the ADM Hamiltonian (Equation~\ref{eq:ADM}). In this section, we compare the analytical results from Sections~\ref{sec:fix2}--\ref{sec:sig} with a numerical analysis that includes the 3.5~PN fluxes. In order to achieve this, we performed numerous numerical integrations of the full 3.5~PN evolution equations for the orbital parameters \citep[see, e.g.][]{mico} and studied the effect of varying the total mass, mass ratio, and initial orbital parameters. Similarly to before, we express the results as a ratio between time-scales. This time, we divide the result from the full 3.5~PN evolution equations with the result from the full 2.5~PN evolution equations (i.e. integrating Equations \ref{eq:at} and \ref{eq:et}). As shown in Figure~\ref{fig:pnfit}, we find that this ratio strongly depends on both orbital parameters -- initial periapsis and eccentricity. The effects of varying total mass and mass ratio are noticeable, but are significantly smaller than our estimation of how much PN tail effects will change the results. Figure~\ref{fig:pnfit} also shows our proposal for a convenient fit to the integrated data. Led by the findings of previous sections, we chose a fitting function of the form $Q_{\rm f} = \exp (C r_{\rm S}/p_0)$ and found the best-fitting value $C = 2.5$, such that

\begin{align}
    Q_{\rm f}(p_0) = \exp \left(2.5 \frac{r_{\rm S}}{p_0} \right). \label{eq:Qf} \\ \nonumber
\end{align}

Similarly to the previous fitting function $R$, the goal of using the ratio $Q_{\rm f}$ is not to perfectly reproduce the full first-order results, but rather to be \textit{sufficiently precise and simple to extend the validity of Peters' formula to more relativistic orbits}. Furthermore, a quick estimation of the effect of PN tail contributions (as shown by the representative error bars in the top panel of Figure~\ref{fig:pnfit}) shows that trying to fit the eccentricity dependence more precisely will not necessarily lead to a better estimation of the true, fully relativistic result. In other words, the simple fitted ratio in Equation~\eqref{eq:Qf} is accurate within 1~PN precision. It can successfully describe the trend of including PN contributions to the orbital dynamics and to the energy fluxes. It is not by chance that this fitted ratio has a very similar form as the analytical correction factor $Q$ found in Sections~\ref{sec:fix2}--\ref{sec:sig}. Indeed, if we perform the PN expansion in the small parameter $r_{\rm S}/p_0$,

\begin{align}
    Q_{\rm f}(p_0) =  \exp \left(2.5 \frac{r_{\rm S}}{p_0} \right) \approx 1+ 2.5\frac{r_{\rm S}}{p_0} + \mathcal{O}(c^{-3}),\\ \nonumber
\end{align}

\noindent we recover the same functional form of $Q = 1 + 5 r_{\rm S}/p_0$.

Even though the coefficient of the fitted ratio is smaller by a factor of two, a simple estimate of the effect of PN tail contributions shows that the uncertainties in the coefficients can overlap. We crudely model the effect of PN tail contributions as follows: if the 1~PN correction to the ratio between time-scales is of the form

\begin{align}
    1+D  \frac{r_{\rm S}}{p_0},\\ \nonumber
\end{align}

\noindent where $D$ is a constant (or a function of eccentricity), then the uncertainty due to the tail correction should be of relative 1.5 PN order. Roughly, we expect it to have the following form:

\begin{align}
    1+D  \frac{r_{\rm S}}{p_0} \pm   \left(D\frac{r_{\rm S}}{p_0}\right)^{\frac{3}{2}},\\ \nonumber
\end{align}

\noindent which can be expressed as some uncertainty in the coefficient $D$:

\begin{align}
    D \to D \pm D^{\frac{3}{2}} \sqrt{\frac{r_{\rm S}}{p_0}}.\\ \nonumber
\end{align}

In this estimation, the uncertainties in the coefficients of the ratios $Q$ and $Q_{\rm f}$ overlap for all orbits with (initial) periapses smaller than approximately $35 r_{\rm S}$ (where the first PN correction is only $\sim$10 per cent). In this region of parameter space, both ratios are, in a sense, equivalent and may be used interchangeably. Nonetheless, we believe the fitted ratio $Q_{\rm f} = \exp(2.5 r_{\rm S}/p_0)$ to be more robust than the analytical ratio $Q$, since it has been produced with more accurate numerical methods. It can be thought of as a ``calibration'' of the analytical result with the inclusion of the 3.5~PN fluxes. We therefore propose it as a simple way to model the effects of the first-order PN perturbations on the time-scale of GW induced decay.

\section{Discussion and Conclusions}\label{sec:conclusions}

In this work, we introduced the two simple factors $R(e_0)$ and $Q_{\rm f}(a_0,e_0)$ that, if multiplied by the standard Peters' formula, can (i) correct for the evolution of the eccentricity and (ii) model the effects of the first-order PN perturbation. The full expression for the corrected Peters' formula is given in terms of the initial semimajor axis $a_0$ and eccentricity $e_0$:

\begin{align}
     t_{\rm P}RQ_{\rm f} = \label{eq:corrpet} 
     \underbrace{ \frac{5c^5(1 + q)^2}{256 G^3 M^3q} \frac{a_0^4}{f(e_0)} }_{\text{Peters' formula}}
    \underbrace{ 8^{ 1- \sqrt{1-e_0}} \exp \left( \frac{2.5 r_{\rm S}}{a_0(1-e_0)} \right)}_{\text{eccentricity and PN correction}}.
\end{align}

\begin{figure}
    \centering
    \includegraphics[scale=0.75]{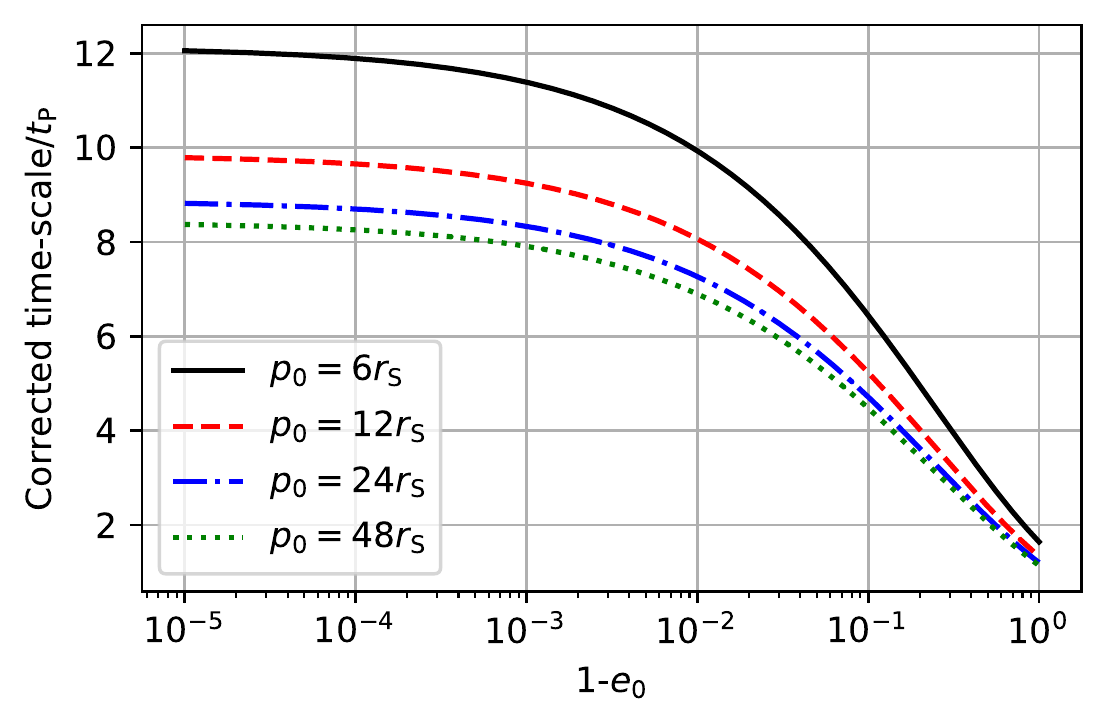}
    \caption{We show the combination of the correction factors $R$ and $Q_{\rm f}$ as a function of initial eccentricity for different initial periapses. Even though the PN correction is only of fractional order by itself, it can compound with the eccentricity-correction factor for very eccentric orbits. GW sources such as EMRIs are expected to originate from precisely these regions of parameter space (high eccentricity and low periapsis). For such sources, using the corrected Peters' formula (Equation~\ref{eq:corrpet}) can resolve errors of up to one order of magnitude in the time-scale of GW induced decay, without needing to resort to numerical integration.}
    \label{fig:QR}
\end{figure}

Figure~\ref{fig:QR} shows the interplay between the PN correction and the eccentricity-evolution correction. Even though the former is only a small fraction, the two factors can compound on each other and produce a correction of more than a factor of ten for very eccentric orbits. This is a clear indication that the standard Peters' time-scale is an insufficient description of the time-scale of GW induced decay for many GW sources.

\begin{figure}
    \centering
    \includegraphics[scale=0.82]{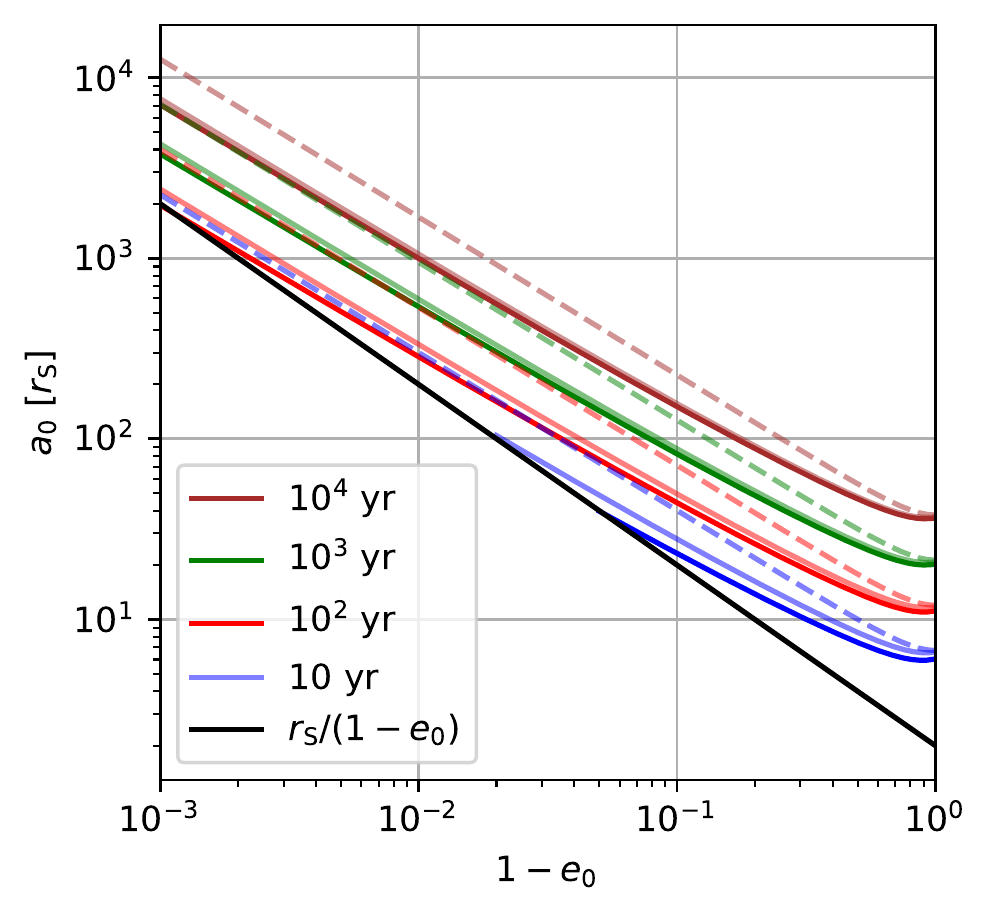}
    \caption{We show a region of phase space in the initial semimajor axis and eccentricity. The panel shows an EMRI (with $m_1 = 10^6$~M$_{\odot}$ and $m_2 = 10$~M$_{\odot}$). The plots would be qualitatively similar for other types of GW sources, such as SMBH binaries and stellar-mass binaries. The solid coloured lines represent the isochrone curves (for different $\tau$) obtained with the corrected Peters' formula ($RQ_{\rm f}t_{\rm P} = \tau$), whereas the dashed coloured lines are obtained with Peters' time-scale ($t_{\rm P} = \tau$). The pale, solid couloured lines are computed with only the eccentricity-evolution correction $R$ (i.e. $Rt_{\rm P} = \tau$). The black solid curve represents orbits whose periapsis is one Schwarzschild radius.}
    \label{fig:isochrones}
\end{figure}

A useful way to visualise the effects of the correction factors $R$ and $Q_{\rm f}$ is to compute isochrone curves. This amounts to setting the time-scale of gravitational decay equal to a fixed time $\tau$ (i.e. $RQ_{\rm f}t_{\rm P} = \tau$ or $t_{\rm P} = \tau$) and, for any given eccentricity, solving for the semimajor axis. We compute the corrected isochrones for the case of an EMRI and plot the results in the ($a_0$, $1-e_0$) phase space in Figure~\ref{fig:isochrones}. The plots would be qualitatively similar for other types of GW sources, such as SMBH binaries and stellar-mass binaries. The effect of the correction factor is to shift the isochrones towards the lower-left corner of the phase-space plots. Equivalently, any point in phase space that used to lay on an isochrone labelled by the time $t_{\rm P}$ (given by Peters' time-scale) now lays on an isochrone labelled with the corrected time $RQ_{\rm f}t_{\rm P}$.

It is worth noting that the discrepancy between the corrected and classical Peters' formula is enhanced for binaries with a large eccentricity. A wide variety of physical processes are known to enhance the eccentricity of a candidate GW source (and thus promote the inspiral) at all mass scales. For instance, the eccentricity of stellar compact binaries may  increase as a result of the supernova kicks experienced by stellar BHs and neutron stars at birth \citep{Giacobbo2019, Giacobbo2019b}; in addition, evolution in triplets can induce large eccentricities due to Kozai--Lidov  \citep{Kozai1962, Lidov1962} cycles and chaotic evolution for both stellar \citep[e.g.][]{Antonini2016,Bonetti_et_al_2018} and SMBH binaries \citep[e.g.][]{Bonetti2016}. As a notable example, \citet{Bonetti2019} showed that evolution in triplets can produce SMBH binaries whose eccentricity is greater than 0.9 when the source enters the LISA band. Finally, most EMRIs are expected to enter the GW phase via the classical two-body relaxation channel \citep{hb,sigurd} with eccentricities very close to unity \citep{Amaro-Seoane2007}. For these reasons, we claim that the correction factors for Peters' time-scale  will have an effect on the predicted detection rates of the LIGO-Virgo and LISA observatories. Peters' time-scale is commonly used to estimate the likelihood of a population of compact objects to produce gravitational signals detectable by GW observatories. It is used both in the modelling of the astrophysics \citep{barausse,oleary} and in signal analysis \citep{klein}. A fractional increase in the time-scale of gravitational decay corresponds to a fractional increase in the time that any signal remains in the optimal frequency band of the observatory. Moreover, it will have some astrophysical implication on the rates at which sources of GWs are produced. The interplay between these two effects will affect how many and what potential sources of GWs are considered promising for both LISA and LIGO-Virgo observatories. We are currently working on a quantitative formulation of this effect, with a focus on the upcoming detector LISA. Preliminary calculations show that it might increase the predicted rates because of the better signal to noise ratios arising from longer decay durations. However, this effect must be weighted against other astrophysical implications of an increased GW time-scale.

As opposed to the correction factors $Q_{\rm f}$ and $R$, the absolute difference between Peters' and the corrected time-scale strongly depends on the mass ratio and total mass of the system. For the special case of EMRIs, mere fractional corrections to the GW time-scale can be of the order of astrophysical time-scales describing other dynamical processes around a SMBH. The correction factors $Q_{\rm f}$ and $R$ might therefore have significant astrophysical implications. Most EMRIs are believed to be generated by the scattering of a stellar-mass compact object to an orbit with very low angular momentum \citep{hb,sigurd} via the two-body relaxation process. However, two-body scatterings can also deflect the orbit of an inspiral before it has had time to complete its gravitational decay. In order to understand whether a compact object reaches its GW induced coalescence, one has to compare the time-scale of gravitational decay to the two-body relaxation time-scale, given by

\begin{align}
    t_{\rm r} = \frac{\sqrt{18 \pi^4}}{32 C_{\gamma}}\left(\frac{G M_{\rm SMBH}}{ a}\right)^{3/2}\frac{(1-e^2)a^{\gamma}}{G^2 m_{\rm s} m_{\rm o} N \ln (\frac{M_{\rm SMBH}}{2 m_{\rm s}})},\label{eq:tr}\\ \nonumber
\end{align}

\noindent where $M_{\rm SMBH} = 4.3 \times 10^6$ M$_{\odot}$ is the SMBH mass, $m_{\rm s} = 1$ M$_{\odot}$ is the mass of a typical field star, $ m_{\rm o}= 10$ M$_{\odot}$ is the mass of the compact object undergoing the process, $N= 10^6$ is the number of stars in a given unit sphere of radius of 1 pc, $\gamma$ is the exponent of the number density profile, and $C_{\gamma}$ is a constant of order unity. The values of the parameters introduced above are chosen to represent the relaxation process in a Milky Way-analogue \citep[see, e.g.][]{elisa}.

\begin{figure}
    \centering
    \includegraphics[scale=0.82]{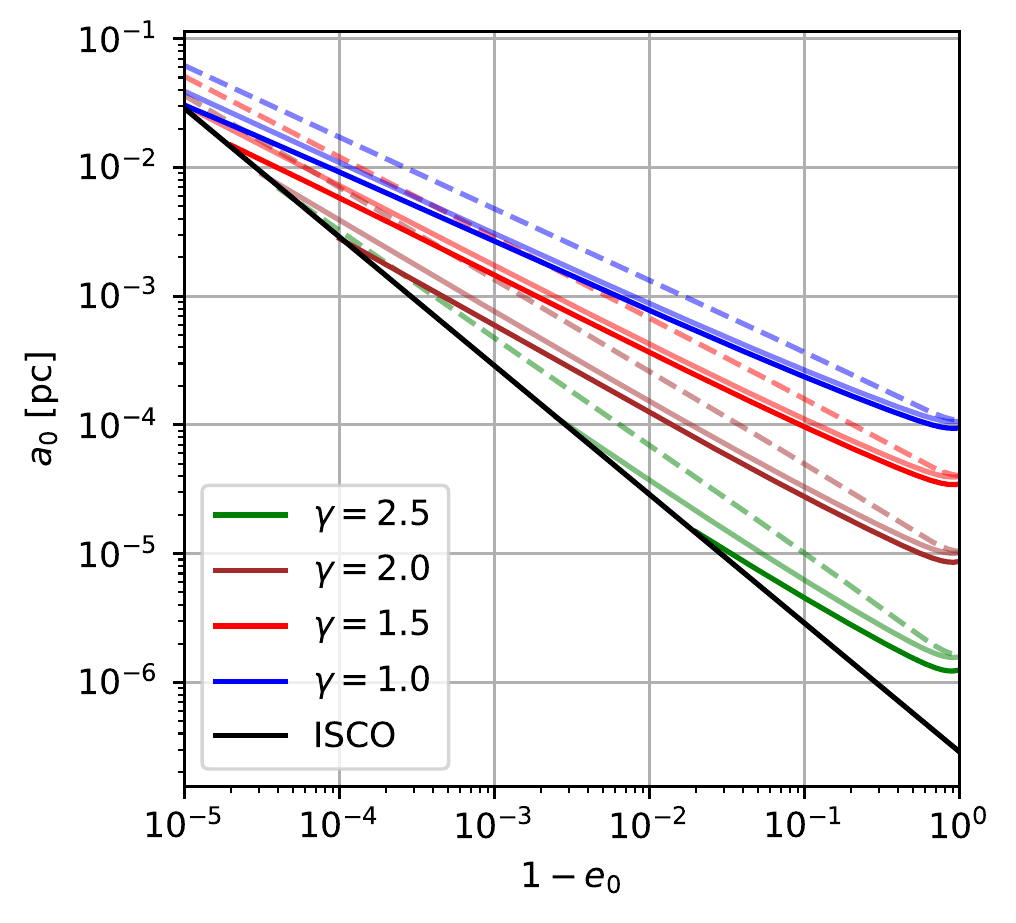}
    \caption{We show a region of phase space in the initial semimajor axis and eccentricity, for an EMRI system with $m_1 = 4.3 \times 10^6$ and $m_2 = 10$~M$_\odot$. The black solid line represents the effective innermost stable circular orbit (ISCO; $3 r_{\rm S}$), whereas the coloured solid (dashed) lines are the curves obtained by equating the two-body relaxation time-scale to the corrected (uncorrected) Peters' time-scale, for four different density power laws ($\gamma =1$; 1.5; 2; 2.5). The pale, solid coloured lines are computed by only taking into account the eccentricity-evolution correction $R$ ($R t_{\rm P} = t_{\rm r}$). Note how the corrected lines ($R Q_{\rm f} t_{\rm P} = t_{\rm r}$) diverge from the uncorrected ones ($t_{\rm P} = t_{\rm r}$) for eccentric orbits, where the PN correction compounds with the eccentricity-evolution correction. Orbits below the ISCO line are expected to decay too quickly to produce easily detectable GW signals and are therefore not relevant. Similarly, orbits above the coloured lines are scattered before gravitational radiation has time to significantly affect their orbit. This leaves us with the phase-space volume between the black and coloured lines as a proxy of the region of potential EMRI candidates. Such volume is drastically reduced by the correction(s) to Peters' time-scale.}
    \label{fig:phase}
\end{figure}

In Figure~\ref{fig:phase}, we plot the curves that equate the two-body relaxation time-scale to the time-scale of gravitational decay for different power-law stellar distributions, solving $t_{\rm P} = t_{\rm r}$ or $R Q_{\rm f} t_{\rm P} = t_{\rm r}$ for the semimajor axis, for a given eccentricity \citep[][]{pau2,merritt,elisa}. In very rough terms, objects that fall above such curves in ($a_0,1-e_0$) phase space are unlikely to undergo an EMRI as their orbit will be likely scattered in a `safer' zone by two-body relaxation.

The effect of the correction factors is again to shift the curves towards the lower-left corner of the phase-space plots.  This means that slices of parameter space that would have previously been considered promising for EMRIs are now removed. For very steep power laws (i.e. $\gamma = 2; \; 2.5$), the effect of the PN correction is very noticeable even for low-eccentricity orbits. Steep density profiles could be produced by the effect of strong mass segregation \citep[][]{Alexander2009, Preto2010} aided by the slow natal kicks received by stellar BHs at birth \citep[][]{Bortolas2017}. For shallower power laws (i.e. $\gamma = 1; \; 1.5$) expected in the case of weak mass segregation \citep{Bahcall1976, Bahcall1977}, the shift is significant for eccentric orbits. Even though preliminary, these plots suggest that the correction factors $Q_{\rm f}$ and $R$ will have a direct effect on previously computed event rates for EMRIs that use Peters' time-scale to model gravitational radiation \citep{pau1,pau2,Babak}. We are currently working on a quantitative estimate on the production rate that takes this into account, along with the associated change in expected EMRI detection rates for LISA.

\subsection{Conclusion}

In this paper, we derive a revised form of Peters' time-scale for the GW induced decay of compact objects. Our key result is represented by the correction factors $R=8^{ 1- \sqrt{1-e_0}}$ and $Q_{\rm f} = \exp(2.5r_{\rm S}/p_0)$, where $e_0$ and $p_0$ are the orbital periapsis and eccentricity, and $r_{\rm S}$ is the effective Schwarzschild radius of the system. The well-known Peters' time-scale must be multiplied by these two factors if one wishes to obtain a more accurate estimate of the duration of a GW induced inspiral, valid for any kind of binary source. We show that the corrected time-scale can reproduce the effects of the self-consistent evolution of the eccentricity as well as model the first-order PN perturbations. The ratio between the uncorrected and the corrected time-scales depends solely on the initial eccentricity and periapsis of the orbit in question and can be of significant magnitude ($RQ_{\rm f} \approx 1$--10) for a large range of orbital parameters. We claim that this difference has an effect on current event-rate predictions for GW sources, as it influences both the astrophysical modelling of gravitational radiation and the quality of detectable signals. With regards to the PN correction factor $Q_{\rm f}$, it is known that BH spin, which we neglected, plays a significant role in the decay time of binaries. This effect enters the picture at 1.5~PN order and might in principle be comparable to the correction computed in this paper due to the unruly nature of the PN series. However, stochastic spin alignments will tend to suppress the effect in a statistical event-rate description that contains many BHs. We are currently working on extensions to the PN correction factor that take BH spin into account, along with PN tail contributions, for applications that focus on a single SMBH. The advantage of the formula $Q_{\rm f} = \exp(2.5r_{\rm S}/p_0)$ is that it captures the largest deviation of fully relativistic orbits from Keplerian ones while remaining algebraically simple. Together with the eccentricity-correction factor $R$, it can replace the Newtonian model currently used for event-rate estimates without complicating the computations or resorting to numerical methods where previously none were employed. Moreover, the corrected formula can function as a computationally efficient replacement for the evolution Equations~\eqref{eq:at} and \eqref{eq:et} -- and their 1~PN extensions -- for applications where such integrations are required. Therefore, we propose that the corrected Peters' formula $RQ_{\rm f}t_{\rm P}$ should be implemented in future event- and detection-rate estimates for LISA and LIGO-Virgo sources, in order to model the effects of gravitational radiation more accurately.

\section*{Acknowledgements}

The authors thank the anonymous reviewer for providing feedback that greatly improved this work. EB, PRC, LM, and LZ acknowledge support from the Swiss National Science Foundation under the Grant 200020\_178949 and thank Matteo Bonetti and Alberto Sesana for fruitful discussions. PAS acknowledges support from the Ram{\'o}n y Cajal Programme of the Ministry of Economy, Industry and Competitiveness of Spain, the COST Action GWverse CA16104, the National Key R\&D Program of China (2016YFA0400702) and the National Science Foundation of China (11721303).

\scalefont{0.94}
\setlength{\bibhang}{1.6em}
\setlength\labelwidth{0.0em}
\bibliographystyle{mnras}
\bibliography{references}
\normalsize

\appendix

\section{Orbits with identical initial energy and angular momentum}\label{sec:fix1}

We start by recalling the explicit first-order expressions for the semimajor axis, $a_{1}$, and eccentricity, $e_{1}$, of the perturbed orbit in terms of the PN conserved energy $E_1$ and angular momentum $h_1$ \citep{QK_Parametrisation}, given by Equations~\eqref{eq:2} and \eqref{eq:2B}.

Since the underlying assumption of this section is that the initial energy and angular momentum in the Newtonian and PN framework are the same, we can replace $E_1$ and $h_1$ with their classical definitions ($E_{\rm N}$ and $h_{\rm N}$):

\begin{align}
    E_1 \to E_{\rm N} &= -\frac{G M }{2 a_{\rm N}}, \label{eq:perturb_a}\\
    h_1^2 \to h_{\rm N}^2 &= \frac{a_{\rm N} \left( 1- e_{\rm N}^2 \right)}{G M}.\\ \nonumber
\end{align}

A useful product of this manipulation is the formula for the perturbed semimajor axis in terms of its Newtonian equivalent, which shows how the PN perturbations slightly reduce the orbit's size:

\begin{align}
    a_1 = a_{\rm N} -\frac{G M (7 + 13 q + 7 q^2)}{4c^2(1 + q)^2}. \label{eq:1}\\ \nonumber
\end{align}

In order to quantify the time-scale for the perturbed orbit's decay, we must manipulate Equation~\eqref{eq:1} to obtain a new variable $a_{\rm m}$, defined as the value of the Newtonian semimajor axis when the binary described via the perturbed orbit can be said to have reached its coalescence. In other words, we seek the value of $a_{\rm N}$ for which the perturbed periapsis $p_1 = a_1(1-e_1)$ of the orbit has reached the effective Schwarzschild radius of the two-body system (hereafter, simply Schwarzschild radius), $r_{\rm S} = 2GMc^{-2}$. If the periapsis is not smaller than the Schwarzschild radius at the beginning of the evolution, an orbit is expected to circularise via the emission of GWs before coalescing (in absence of other dynamical effects), allowing us to replace $p_1 = a_1(1-e_1)$ with $a_1$ in the following equation:

\begin{align}
 p_1 = a_1 \overset{!}{=} \frac{2 G M}{c^2} \implies  a_{\rm m} = \frac{G M \left(15 q ^2+29 q +15\right)}{4 c^2 (1+q )^2}. \label{eq:e}\\ \nonumber
\end{align}

In order to compare the time-scales for the decay, we can simply integrate Equation~\eqref{eq:at} in the two different cases: the Newtonian orbit has to decay until the Schwarzschild radius is reached, whereas the perturbed orbit has to decay until the value $a_{\rm m}$ is reached. Since $a_{\rm m}$ is always larger than the Schwarzschild radius, we come to a qualitative result: \textit{PN orbits decay more quickly than what Peters' formula predicts, when comparing binaries with identical initial energy and angular momentum.}

In order to solve the decay Equation~\eqref{eq:at} analytically, we have to make the simplifying assumption of small initial eccentricity ($e_{\rm N}^2 \approx 0$). For a given initial semimajor axis $a_0$, the time evolution is given by Equation~\eqref{eq:3}.

The decay process takes a binary from its initial semimajor axis $a_0$ (and associated energy) to coalescence. To know the duration of this process, we set Equation~\eqref{eq:3} equal to the value of the last allowed semimajor axis before the coalescence is achieved, and solve for the variable $t$. We obtain two different time-scales -- $t_{\rm P}$ for a Newtonian orbit (Peters' time-scale) and $t_{\rm c}$ for a perturbed orbit (corrected time-scale) -- by respectively choosing the final semimajor axis values as $2G M c^{-2}$ and $a_{\rm m}$. To improve readability, we express the time-scales in terms of multiples of the Schwarzschild radius, by defining the number $n \ge 1$ via  $n = a_0/r_{\rm S} $. We obtain

\begin{align}
    t_{\rm P} &= \frac{5 \left(4 n^4-1\right) (q +1)^2 r_{\rm S}}{128 c q },\\
    t_{\rm c} &=\frac{5  \left(16384 n^4 (q +1)^8-\left(15 q ^2+29 q +15\right)^4\right)r_{\rm S}}{524288 c
   q  (q +1)^6}.\\ \nonumber
\end{align}

The simplest way to account for large initial eccentricities is to note that most of the decay time is spent in the neighbourhood of the initial conditions \citep[this is the same argument used in][]{Peters_1964}. Therefore, we can adjust the time-scales by dividing them by the eccentricity enhancement function $f(e)$ evaluated at the initial eccentricity:

\begin{align}
    t_{\rm P} \to \frac{t_{\rm P}}{f(e_0)},\;\;\;\;\; t_{\rm c} \to \frac{t_{\rm c}}{f(e_0)}.\\ \nonumber
\end{align}

This simplification allows us to neglect the time evolution of the eccentricity. It is widely used in the literature, even though it can produce large deviations from the self-consistently integrated value of the decay time. Here, we analyse the simpler case of no eccentricity evolution, which serves as a lower bound for the self-consistently evolved decay time. The equations listed so far display a dependence on the binary mass ratio. For convenience, we report here the ratio $\nu_{\rm E} = t_{\rm c}/t_{\rm P}$ and the difference $\delta_{\rm E} = t_{\rm c}-t_{\rm P}$ in the two limiting cases of $q \to 0$,

\begin{align}
    \nu_{\rm E} &=\frac{4 \left(n^4-\frac{50625}{16384}\right)}{4 n^4-1} \le 1, \label{eq:ratioE}\\
    \delta_{\rm E} &= -\frac{232645}{524288} \frac{r_{\rm S}}{cq  }\frac 1 { f(e_0) } \le 0. \label{eq:differenceE}\\ \nonumber
\end{align}

\noindent and $q \to 1$,

\begin{align}
     \nu_{\rm E} &=\frac{4 \left(n^4-\frac{12117361}{4194304}\right)}{4 n^4-1} \le 1, \label{eq:ratioElargeq}\\
    \delta_{\rm E} &= -\frac{55343925 }{33554432} \frac{r_{\rm S}}{cq  }\frac 1 { f(e_0) } \le 0, \label{eq:differenceElargeq} \\\nonumber
\end{align}

\noindent for orbits whose initial energy and angular momentum are the same. The result that PN orbits decay more rapidly when starting with the same initial conditions in energy and angular momentum reflects well known findings from direct integrations of the Hamiltonian.

\bsp 
\label{lastpage}
\end{document}